\documentclass[oldversion]{aa}
\usepackage{amsmath,graphics,amssymb}

\usepackage{times}
\usepackage{natbib}
\bibpunct{(}{)}{,}{a}{}{,}

\newcommand{\Teff}{\ensuremath{T_\mathrm{eff}}}
\newcommand{\zav}[1]{\left(#1\right)}

\newcommand{\hzav}[1]{\left[#1\right]}

\newcommand\sigorie{$\sigma$~Ori~E}

\allowdisplaybreaks

\begin{document}

\title{Mass loss in main-sequence B stars}

\author{Ji\v r\'\i\ Krti\v{c}ka}
\authorrunning{J. Krti\v{c}ka}

\institute{\'Ustav teoretick\'e fyziky a astrofyziky, Masarykova univerzita,
            CZ-611 37 Brno, Czech Republic, \email{krticka@physics.muni.cz}}

\date{Received}

\abstract{We calculate radiatively driven wind models of main-sequence B stars
and provide the wind mass-loss rates and terminal velocities. The main-sequence
mass-loss rate strongly depends on the stellar effective temperature. For the
hottest B stars the mass-loss rate amounts to
$10^{-9}\,\text{M}_\odot\,\text{year}^{-1}$, while for the cooler ones the
mass-loss rate is lower by more than three orders of magnitude. Main-sequence B
stars with solar abundance and effective temperatures lower than about
$15\,000\,\text{K}$ (later than spectral type B5) do not have any
homogeneous line-driven wind. We predict the wind mass-loss rates for the solar
chemical composition and for the modified abundance of heavier elements to study
the winds of chemically peculiar stars. The mass-loss rate may either increase or
decrease with increasing abundance, depending on the importance of the
induced emergent flux redistribution. Stars with overabundant silicon may have
homogeneous winds even below the solar abundance wind limit at
$15\,000\,\text{K}$. The winds of main-sequence B stars lie below the
static limit, that is, a static atmosphere solution is also possible. This
points to an important problem regarding the initiation of these winds.
We discuss the implications of our models for 
rotational braking, filling the magnetosphere of Bp stars, and for chemically
peculiar stars.

}

\maketitle

\section{Introduction}

Line-driven winds are most often studied in the most luminous hot stars. These
stars have strong winds that are easy to observe and that
significantly influence the stellar evolution. However, winds of less
luminous stars, especially main-sequence B stars, also have considerable
observational and theoretical consequences. Although the winds in these stars
do not manifest themselves as prominent spectral features or have a
decisive influence on the way the stars evolve, weaker winds affect a number
of observational phenomena that gain considerable astrophysical attention.

Main-sequence B stars lie at the boundary between luminous O stars, where the
radiative force expels huge amounts of mass from the star \citep{pulvina},
and main-sequence A stars, where the minute effect of the radiative levitation
horizontally redistributes the elements in easily recognizable surface spots
\citep{mpoprad}. It has long been speculated that the outflow is a missing
ingredient that is needed to quantitatively explan the resulting abundance
anomalies. Such an explanation faces a strong problem because we lack consistent
theoretical predictions of wind mass-loss rates at low luminosity.

Traditionally, two parameters are invoked to explain all aspects of the
evolution of single stars: the initial mass and the chemical composition. However,
it now seems that at least a third parameter is essential to explain 
the single-star evolution: the initial stellar angular momentum \citep{meem}.
Consequently, rotation becomes one of the crucial parameters of evolution
models. The stellar rotation may not only contribute to the internal mixing of
chemical elements, but it may also bring a star near critical
rotation \citep{sylsit}. Fast rotation is prerequisite for the collapsar
model of gamma-ray bursts \citep{yooni}. Unfortunately, the evolutionary time
scale is so long that testing the predictions of rotational velocity
evolution models directly is a forbidding task. However, there is one exception: the rotationally
modulated light variations of chemically peculiar stars provide a powerful
clock that can reveal even minute changes of the rotational period.
Studies based on such a clock \citep{pyper98,zmenper2} have shown period
variations that might be connected with angular-momentum loss via magnetized
stellar winds. However, precise mass-loss rate predictions for these stars are
lacking, therefore one cannot test the angular momentum loss hypothesis quantitatively.

As a result of its low density, the wind material might be trapped in the
magnetospheres of Bp stars. Resulting magnetospheric clouds are the prime origin
of the light variability of \sigorie\ \citep{labor,towog}. Detailed properties
of these clouds, such as the time scale of their build-up, are poorly constrained
due to uncertain mass-loss rates at low luminosity.

Last but not least, winds of low-luminosity hot stars are also appealing
theoretically. They provide a unique environment in which many astrophysical
phenomena can be studied, including the particle collisions that lead to the
multicomponent nature of the flow \citep[e.g.,][]{treni,kkiv}, or the line
Doppler-heating that influences the wind temperature \citep{go}. Note, however, that
because the latter two effects occur in the outer wind, they typically do not
affect the mass-loss rate.

Stellar winds of main-sequence B stars are difficult to study observationally.
The H$\alpha$ line, which is a crucial mass-loss-rate indicator in O stars, is
unsuitable for determining a mass-loss rate of the order of
$10^{-8}\,\text{M}_\odot\,\text{year}^{-1}$ or lower \citep{pulvina}, that is, in
the domain of main-sequence B stars. Moreover, the mass-loss rate determination
of main-sequence B stars from observations may be uncertain not only as the
result of the effect of clumping on the line profiles \citep{sund,clres1}, but
also as the result of the weak-wind problem
\citep[e.g.,][]{bourak,martclump,linymarko}. The latter problem is connected
with observed ultraviolet wind-line profiles in low-luminosity O stars, which
are weaker than theory predicts. The weak mass-loss rates inferred from weak
ultraviolet lines are (at least in one case) supported by the mass-loss rate
derived from the Br$\gamma$ line \citep{nabrg}.

On the other hand, there are independent observational indications
\citep{hueneco,gvaneco} that show that the ultraviolet line profiles significantly
underestimate the mass-loss rate in low-luminosity O stars. The weakness of the
ultraviolet line profiles can be explained by the high wind temperature as
a result of the shock heating in a medium with such a low density that is not
able to cool down radiatively \citep{luciebila,cobecru,nlteiii,lucyjakomy}. From
this point of view the theoretical models provide a reliable choice for
estimating B star mass-loss rates.

To improve the situation of poorly known wind parameters in main-sequence B
stars, we here provide mass-loss rate estimates based on our own NLTE wind code.

\section{Description of models}

To predict main-sequence B star mass-loss rates we applied the NLTE wind
models of \citet{cmf1} with a comoving frame (CMF) line force. Our wind
models assume stationary and spherically symmetric flow. They enable us to
self-consistently predict wind structure and wind parameters just from the basic
stellar parameters, that is, the effective temperature, mass, radius, and chemical
composition.

The ionization and excitation state of considered elements was derived from the
statistical equilibrium (NLTE) equations. For this purpose we adopted a set of
ionic models from the TLUSTY grid of model stellar atmospheres
\citep{ostar2003,bstar2006}, which was extended using the data from the Opacity
and Iron Projects \citep{topt,zel0}. For phosphorus we employed data described
by Pauldrach et al. (\citeyear{pahole}).

The NLTE model atmosphere emergent fluxes, which serve as a lower boundary
condition of the radiative transfer equation in the wind, are calculated by
the TLUSTY code \citep{bstar2006} for the same stellar parameters (the effective
temperature, surface gravity, and chemical composition) as the wind models.

The line radiative force was calculated using the solution of the spherically
symmetric CMF radiative transfer equation \citep{mikuh}. The corresponding line
data were extracted from the VALD database (Piskunov et al. \citeyear{vald1},
Kupka et al. \citeyear{vald2}). We applied the electron thermal balance method
(Kub\'at et al.~\citeyear{kpp}) to calculate the radiative cooling and
heating. These terms were calculated with the occupation numbers derived from the
statistical equilibrium equations.

The continuity equation, equation of
motion, and energy equation were solved iteratively
using the Newton-Raphson method. An accelerating outflow was assumed in the
first step of iterations. As a result of iterations we  obtained the wind density,
velocity, and temperature structure that satisfy the hydrodynamical equations. We
found no case where the derived solution (faster than
the radiative-acoustic waves at outer edge, \citealt{abb}) was not unique, that
is, the
final converged solution does not depend on a particular choice of initial
conditions. We assumed homogeneous flow, that is, we
neglect possible differences in velocities of individual components. This is
legitimate here, because we are mainly interested in the mass-loss rates.

\begin{table}
\caption{Stellar parameters of the model grid}
\centering
\label{bhvezpar}
\begin{tabular}{ccccccc}
\hline
Model & Sp. & $\Teff$ & $M$ & $R_{*}$ & $\log(L/L_\odot)$ & 
$v_\text{esc}$ \\
&Type&$[\text{K}]$ & $[{M}_{\odot}]$ & $[{R}_{\odot}]$  & &
[$\text{km}\,\text{s}^{-1}$]\\
\hline
T30 & B0 & 30\,000 & 14.73 & 5.84 & 4.39 & 960\\
T28 &    & 28\,000 & 12.69 & 5.34 & 4.20 & 940\\
T26 & B1 & 26\,000 & 10.88 & 4.87 & 3.99 & 910\\
T24 &    & 24\,000 &  9.26 & 4.45 & 3.77 & 880\\
T22 &    & 22\,000 &  7.84 & 4.07 & 3.54 & 850\\
T20 &    & 20\,000 &  6.60 & 3.71 & 3.30 & 820\\
T18 &    & 18\,000 &  5.52 & 3.39 & 3.03 & 790\\
T16 & B5 & 16\,000 &  4.58 & 3.09 & 2.75 & 750\\
T14 & B6 & 14\,000 &  3.76 & 2.80 & 2.43 & 720\\
\hline
\end{tabular}
\end{table} 

The stellar parameters of the considered model grid are given in
Table~\ref{bhvezpar}. Here we also provide the corresponding spectral types for
selected models. The effective temperature $\Teff$ covers the range of earlier B
stars, and the stellar mass $M$ and radius $R_{*}$ were calculated from the
interpolation formulas derived for main-sequence stars by \citet{har} with the
effective temperature as a parameter. To calculate both wind and
atmosphere models we assumed the solar chemical composition of \citet{asp09}. We
also calculated additional models with a modified abundance of the elements that
are most important for the line driving, that is, carbon, nitrogen, oxygen,
silicon, sulfur, and iron. The range of considered abundances of these elements
was partially motivated by the abundances typically found on the surface of chemically
peculiar stars \citep[e.g.,][]{briketka,leh2,bohacen}.

\section{Mass-loss rates}

\begin{figure}[t]
\centering \resizebox{\hsize}{!}{\includegraphics{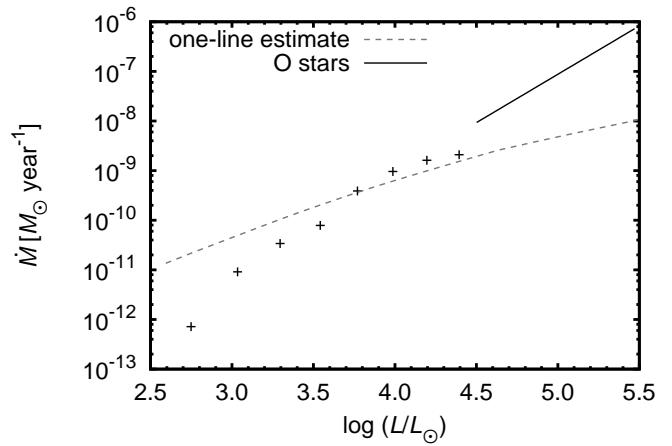}}
\caption{Predicted mass-loss rates for solar metallicity as a function of
the stellar luminosity. Overplotted (solid line) is the fit to main-sequence O
star mass-loss rates \citep{fosfor}. The dashed line denotes the one-line
mass-loss rate estimate \citep{cnovit}.}
\label{dmdtl}
\end{figure}

\begin{table*}
\caption{Predicted mass-loss rates and terminal velocities for different
abundances and different effective temperatures. Mass-loss rates are given in
units of $10^{-11} M_\odot\,\text{year}^{-1}$.
The terminal velocities $v_\infty$ are given in units of
$\text{km}\,\text{s}^{-1}$. The terminal velocity in parentheses denotes a model
for which it was not possible to calculate a monotonic solution to large radii
(see the text).}
\centering
\label{hpvys}
\begin{tabular}{cccccccccccccccccccccccccccc}
\hline
&\multicolumn{2}{c}{Solar}&
\multicolumn{2}{c}{$\text{[C/H]}=-1$}&
\multicolumn{2}{c}{$\text{[C/H]}=1$}&
\multicolumn{2}{c}{$\text{[N/H]}=-1$}&
\multicolumn{2}{c}{$\text{[N/H]}=1$}&
\multicolumn{2}{c}{$\text{[O/H]}=-1$}&
\multicolumn{2}{c}{$\text{[O/H]}=1$}\\
Model & 
$\dot M$ & $v_\infty$ &
$\dot M$ & $v_\infty$ &
$\dot M$ & $v_\infty$ &
$\dot M$ & $v_\infty$ &
$\dot M$ & $v_\infty$ &
$\dot M$ & $v_\infty$ &
$\dot M$ & $v_\infty$ \\
\hline
T30 & 210 & 3650& 190 & 2280& 210 & 5490& 120 & 4010& 370 & 4410& 100 & 4110& 190 & 4390\\
T28 & 160 & 4380& 150 & 2620& 110 & 8610& 67 & 4540& 160 & 5520& 160 & 4500& 130 & 4260\\
T26 & 96 & 3070& 60 & 2400& 60 & (2340)& 64 & (2440)& 90 & 4790& 100 & 2990& 79 & 3040\\
T24 & 39 & 1700& 23 & 1720& 42 & 1820& 4.8 & 2040& 38 & 3620& 25 & 1850& 39 & 1700\\
T22 & 7.9 & 1690& 8.5 & 1580& 14 & 1490& 0.71 & 1760& 14 & 2730& 6.5 & 1820& 12 & 1600\\
T20 & 3.4 & 1290& 2.3 & 1250& 1.5 & 1410& 1.6 & 1260& 6.4 & 1720& 3.2 & 1300& 3.6 & 1200\\
T18 & 0.91 & 820& 0.42 & 460& 0.58 & 900& 0.51 & 760& 1.8 & 1120& 0.84 & 720& 0.88 & 790\\
T16 & 0.072 & (240)& 0.023 & (180)& 0.06 & 520& 0.038 & (230)& 0.17 & 420& 0.058 & (240)& 0.06 & (230)\\
T14 &\multicolumn{2}{c}{no wind} &\multicolumn{2}{c}{no wind} &\multicolumn{2}{c}{no wind} &\multicolumn{2}{c}{no wind} &\multicolumn{2}{c}{no wind} &\multicolumn{2}{c}{no wind} &\multicolumn{2}{c}{no wind}\\

\end{tabular}
\begin{tabular}{cccccccccccccccccccccccccccc}
\hline
&\multicolumn{2}{c}{$\text{[Si/H]}=1$}&
\multicolumn{2}{c}{$\text{[Si/H]}=2$}&
\multicolumn{2}{c}{$\text{[Si/H]}=3$}&
\multicolumn{2}{c}{$\text{[S/H]}=-1$}&
\multicolumn{2}{c}{$\text{[S/H]}=1$}&
\multicolumn{2}{c}{$\text{[Fe/H]}=1$}&
\multicolumn{2}{c}{$\text{[Fe/H]}=2$}\\
Model & 
$\dot M$ & $v_\infty$ &
$\dot M$ & $v_\infty$ &
$\dot M$ & $v_\infty$ &
$\dot M$ & $v_\infty$ &
$\dot M$ & $v_\infty$ &
$\dot M$ & $v_\infty$ &
$\dot M$ & $v_\infty$ \\
\hline
T30 & 250 & 4020& 320 & 4490& 490 & 2910& 180 & 3720& 380 & 3390& 230 & 3040& 1700 & 1720\\
T28 & 180 & 5510& 210 & 5180& 240 & 3200& 130 & 4640& 280 & 3850& 140 & 3570& 370 & 1910\\
T26 & 110 & (5270)& 150 & 3720& 160 & 3070& 83 & 3220& 170 & 2820& 83 & 3060& 57 & 2390\\
T24 & 54 & 4580& 81 & (1620)& 77 & 2450& 29 & 1760& 81 & 1710& 31 & 1790& 16 & 1740\\
T22 & 21 & 4020& 23 & 6390& 9.9 & 2250& 5.3 & 1820& 22 & 1730& 6 & 1900& 8.5 & 1690\\
T20 & 7.5 & 3630& 8.4 & 7740& 4.6 & 380& 3 & 1270& 4.5 & 1360& 2.6 & 1320& 1.8 & 1290\\
T18 & 3.1 & 1990& 3.6 & 5270& 1.9 & 510& 0.84 & 790& 1.4 & 830& 0.74 & 830& 0.39 & 790\\
T16 & 0.7 & 1400& 0.7 & 2440& 0.17 & (170)& 0.056 & (240)& 0.085 & (270)& 0.041 & (250) &\multicolumn{2}{c}{no wind}\\
T14 & 0.01 & 650& 0.004 & 1290& 0.002 & 1550 &\multicolumn{2}{c}{no wind} &\multicolumn{2}{c}{no wind} &\multicolumn{2}{c}{no wind} &\multicolumn{2}{c}{no wind}\\

\hline
\end{tabular}
\end{table*}

The predicted mass-loss rates $\dot M$ for B stars with solar metallicity in Table~\ref{hpvys}
nicely follow the O star mass-loss rate -- luminosity relationship
(Fig.~\ref{dmdtl}). The mass-loss rates of solar metallicity B stars can be
fitted via the formula
\begin{equation}
\label{metuje}
\log\zav{\frac{\dot M}{1\,{M}_\odot\,\text{year}^{-1}}}=a+
b\zav{\frac{\Teff}{10^4\,\text{K}}}+c\zav{\frac{\Teff}{10^4\,\text{K}}}^2,
\end{equation}
where
\begin{align}
a=&-22.7, &b=&8.96, &c=&-1.42.
\end{align}
The predicted mass-loss rates can be compared with the mass-loss-rate estimate
corresponding to one optically thick line $\dot M=8\pi^2 R_*^2\nu_{ij}
H_c(\nu_{ij})/c^2$ \citep[Eq.~(11)]{cnovit}. In this formula
$H_c(\nu_{ij})$ is the flux at the line frequency $\nu_{ij}$, but in
Fig.~\ref{dmdtl} we use  the maximum product of flux and frequency, giving the
maximum mass-loss rate with one optically thick line. For B stars with higher
luminosities this one-line mass-loss-rate estimate roughly corresponds to the
predicted mass-loss rate because there are just a few lines that drive the wind.
For stars with lower luminosities the one-line mass-loss-rate estimate
overestimates the mass-loss rate because in these stars even the strongest lines
become optically thin and their positions do not correspond to the
maximum product $\nu H_c(\nu)$.

\begin{table}
\caption{Line force multipliers for the models with solar metallicity}
\centering
\label{cakpar}
\begin{tabular}{cccc}
\hline
Model & $k$ & $\alpha$ &  \vrule height2.3ex width0pt $\bar Q$\\
\hline
T30 &0.024 & 0.635 & 35\\
T28 &0.019 & 0.670 & 54\\
T26 &0.044 & 0.610 & 95\\
T24 &0.38 & 0.435 & 110\\
T22 &0.20 & 0.46 & 61 \\
T20 &0.60 & 0.390 & 95 \\
T18 &2.50 & 0.295 & 130\\
T16 &29.5 &  0.140 & 210\\
\hline
\end{tabular}
\end{table} 

In main-sequence B stars the wind mass-loss rate significantly decreases with
decreasing effective temperature (or luminosity). For a decrease of a factor of two 
in the effective temperature the decrease in mass-loss rates is more than
three orders of magnitude. We were unable to find any wind solution for the solar
metallicity model T14. This indicates that a
homogeneous wind is not possible for $T_\text{eff}\lesssim15\,000\,\text{K}$.

In Table \ref{cakpar} we provide the radiative force multipliers
\citep{cak,abpar} corresponding to our solar-metallicity models. The force
multipliers describe the line distribution function \citep{pusle}, but in our
approach we simply selected force multipliers that provided the best fit of
wind mass-loss rate and terminal velocity derived in models with CMF line force.
The parameter $\alpha$ decreases with the effective temperature to account for
the decrease of the wind terminal velocity. To compensate for this the line
force parameter $k$ increases on average with decreasing effective temperature.
From this point of view the $\bar Q$ parameter of \citet{gayley} is
advantageous, because its variations are much weaker in the considered stars, $\bar
Q\approx100$ (see Table~\ref{cakpar}, where we provide $\bar
Q=\hzav{(1-\alpha)k\zav{c/v_\text{th}}^\alpha}^{1/(1-\alpha)}$, where the
fiducial hydrogen thermal speed is
$v_\text{th}=\sqrt{2kT_\text{eff}/m_\text{H}}$).

There are not many mass-loss rate predictions that can be compared with
Eq.~\eqref{metuje}. \citet{vikolamet} predicted a mass-loss
roughly by a factor 2 higher for the T30 model, but the T30 model is slightly outside the model
grid considered by them. \citet{pahole} predicted a mass-loss rate for the model with
$\Teff=30\,000\,\text{K}$ that is higher by a factor of 4
than that of the T30 model, but for a radius larger by a factor of 2. After
correcting for the dependence of the mass-loss rate on radius \citep[assuming a
rough dependence $\dot M\sim L^2$,][]{fosfor}, this gives a mass-loss rate
that is lower by a factor of $5$. When we extrapolate the rates of
\citet{uncno},
again outside their assumed grid, our predictions are about a factor of $5$
higher for the model T30, while they are higher by about a factor of $3$ for
$\Teff=25\,000\,\text{K}$. Finally, for $\Teff=30\,000\,\text{K}$ our models
predict roughly one eighth of the rates predicted by \citet{jaga}, while for
$\Teff=22\,000\,\text{K}$ our results are nearly the same.

\begin{figure}[t]
\centering \resizebox{\hsize}{!}{\includegraphics{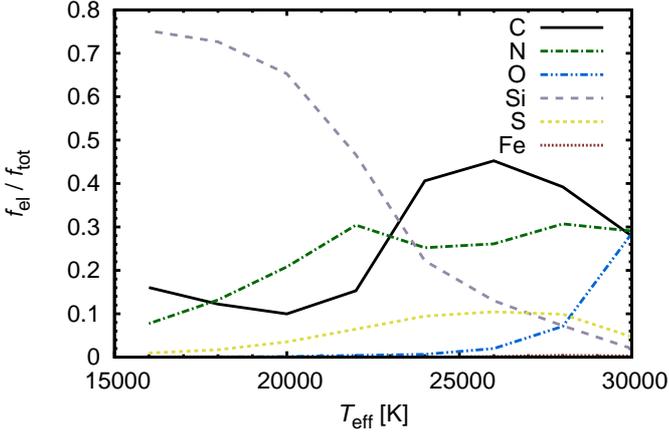}}
\caption{Relative contribution of individual elements to the radiative force
at the critical point as a function of the effective temperature. Models with
solar abundance are plotted here. The contribution of iron is negligible.}
\label{hpsil}
\end{figure}

In Fig.~\ref{hpsil} we plot the relative contribution of individual elements to
the radiative force in models with solar chemical composition. The relative
contribution is plotted at the critical point, where the mass-loss rate of our
models is determined. The contribution of individual elements to the radiative
force is a strong function of the effective temperature. Carbon and nitrogen
dominate the radiative force in the hottest B stars with
$T_\text{eff}\approx30\,000\,\text{K}$, while in stars with
$T_\text{eff}\lesssim22\,000\,\text{K}$ the lines of silicon dominate the
radiative driving.

In contrast to O stars, there are only a few lines that drive the wind in B stars.
Iron, which dominates at higher mass-loss rates \citep[e.g.,][]{vikolamet}, has
a negligible contribution to the radiative force in B stars, because (as a
result of its lower abundance compared to CNO elements) most of the iron lines become
optically thin. However, this is not the only reason for a low
contribution of iron to the radiative force in B stars, because iron has nearly
the same abundance as silicon, which dominates the line driving in cooler B
stars. In contrast to lighter elements such as silicon or carbon, 
iron ions do not have allowed transitions between the most populated ground level and excited
levels with the lowest energies. Consequently, the iron resonance lines have
wavelengths below the Lyman limit, where the stellar radiative flux is too
low at given effective temperatures and does not contribute significantly to the radiative
force.

A similar situation also occurs for oxygen, where the most abundant ion in
B-star wind, \ion{O}{iii}, has the resonance lines in the Lyman continuum. On the
other hand, carbon, nitrogen, and silicon have strong resonance lines in the
Balmer continuum, therefore they significantly contribute to the radiative
force. The importance of silicon is connected with the fact that the \ion{Si}{iv}
resonance lines lie close to the flux maximum. On the other hand, 
carbon is most significant at around $T_\text{eff}=25\,000\,\text{K}$, where the relative
contribution of resonance \ion{C}{iv} lines at 1548\,\AA\ and 1551\,\AA\ is the
strongest, while these lines become optically thin at lower effective
temperatures because of the ionization shift from \ion{C}{iv} to \ion{C}{iii}.

\begin{table}[t]
\caption{Strongest lines in the solar-metallicity models}
\label{carytab}
\begin{tabular}{ll}
\hline
Model & lines (ions and wavelengths in \AA)\\
\hline
T30 & \ion{C}{iii} 977, 1176; \ion{C}{iv} 1548, 1551; \ion{N}{iii} 990, 992;
\\ &\ion{N}{iv} 765, 923; \ion{O}{iii} 833, 834, 835; \ion{O}{iv} 788, 790; \\ &\ion{P}{v} 1118, 1128; 
\ion{S}{v} 786\\ 
T28 & \ion{C}{iii} 977, 1176; \ion{C}{iv} 1548, 1551;  \ion{N}{iii} 990, 992;
\\ &\ion{Si}{iv} 1394, 1403;  \ion{P}{v} 1118; \ion{S}{iv} 1063, 1073\\
T26 & \ion{C}{iii} 977, 1176; \ion{C}{iv} 1548, 1551; \ion{N}{iii} 990, 992;
\\ & \ion{Si}{iv} 1394, 1403; \ion{S}{iv} 1063, 1073\\
T24 & \ion{C}{iii} 977, 1176; \ion{C}{iv} 1548, 1551; \ion{N}{iii} 990, 992;
\\ & \ion{Si}{iv} 1128, 1394, 1403; \ion{S}{iv} 1063, 1073\\
T22 & \ion{C}{iii} 977; \ion{C}{iv} 1548, 1551; \ion{N}{iii} 990, 992;
\\ & \ion{Si}{iv} 1394, 1403; \ion{S}{iv} 1063, 1073\\
T20 & \ion{C}{iii} 977; \ion{N}{iii} 990, 992; \ion{Si}{iv} 1394, 1403; 
\ion{S}{iv} 1073\\
T18 & \ion{C}{iii} 977; \ion{N}{iii} 990, 992; \ion{Si}{iv} 1394, 1403\\
T16 & \ion{C}{iii} 977; \ion{N}{iii} 990, 992; \ion{Si}{iii} 1207; 
\ion{Si}{iv} 1394, 1403\\
\hline
\end{tabular}
\end{table}

These results are also shown in Table~\ref{carytab}, where we list the
strongest lines that drive the wind in the region of the critical point. These lines
can be used as potential wind indicators in B stars. On the other hand, their
absence in the spectra does not mean that the wind does not exist, because we
neglected other potentially important effects, for example, the wind X-ray
ionization in the outer parts of the wind \citep[e.g.,][see also
Sect.~\ref{modelsi}]{oskibp}.

\begin{figure}[t]
\centering \resizebox{\hsize}{!}{\includegraphics{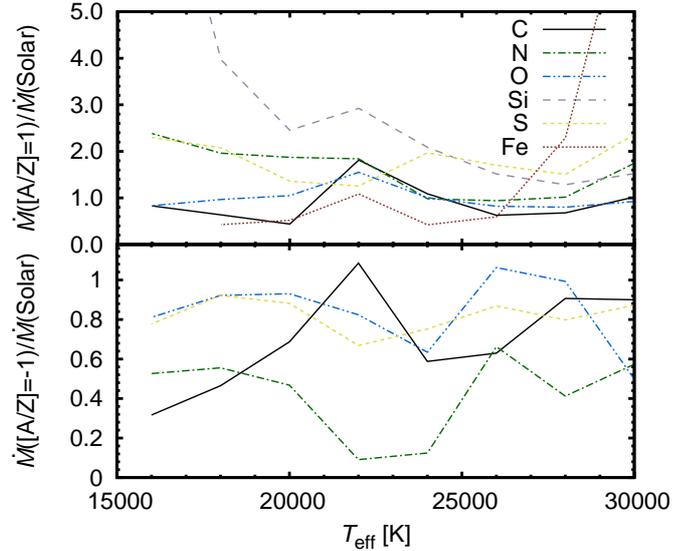}}
\caption{Ratio of the mass-loss rate calculated with an abundance of
a given element enhanced by a factor of ten ({\em upper panel}) and with an 
abundance reduced by a factor of ten ({\em lower panel}) to the mass-loss rate for a
solar chemical composition as a function of stellar effective temperature.}
\label{dmdttz}
\end{figure}

It is common consensus that the mass-loss rate increases with increasing
metallicity because the line absorption inceases. While this is true in
most cases \citep[e.g.,][]{vikolamet,uncno}, Fig.~\ref{dmdttz} shows that this
is not always the case in the domain of main-sequence B stars. In several cases
the mass-loss rate decreases with increasing abundance. This is caused by the
influence of abundance on the emergent flux from the stellar atmosphere.
Typically, the increase of abundance causes the flux redistribution from the far-ultraviolet part of the spectrum (where most of the flux is emitted) to the near
ultraviolet and the visual part. If the redistribution is strong enough, it
may affect the radiative force. We note that the same effect causes the light
variability of chemically peculiar stars \citep[e.g.,][]{peter,molnar}.
Moreover, the stronger atmospheric lines with increasing abundance
lower the line flux and hence the radiative force if these lines significantly
contribute to the radiative driving \citep{babelb}.

The increase of the carbon abundance causes the atmosphere flux redistribution
from the wavelengths shorter than $1100\,$\AA\ to longer wavelengths, which in
turn causes
the decrease of the radiative force. Interestingly, the decrease of the carbon
abundance also causes the decrease of the radiative force as a result of the decrease
of the line absorption. This effect is apparent in models T16-T20 and
T26-T30, whereas it is suppressed by stronger \ion{C}{iv} lines  in
models T22 and T24 with overabundant carbon [C/H]=1.
On the other hand, nitrogen, oxygen, and sulfur do not
significantly influence the emergent atmosphere flux. Consequently, in this case
the radiative force increases with increasing abundance of these
elements. Silicon influences the emergent atmospheric flux for higher abundances
$\text{[Si/H]}\gtrsim2$. As a result, the mass-loss rate only strongly depends on the
silicon abundance when the silicon is not 
significantly overabundant, that is, for $\text{[Si/H]}\lesssim1$. Most interestingly, the
mass-loss rate decreases with increasing iron abundance even if the iron does
not contribute to the radiative force. This is again caused by the redistribution of the
atmosphere radiative flux from the far-ultraviolet region to the
near-ultraviolet and optical regions as a result of flux-blocking by numerous iron
lines. 

The increasing abundance of heavier elements (with the exception of silicon) does
not help to find a wind solution for model T14. This is denoted as ``no wind''
in Table~\ref{hpvys}. Only stars enhanced by silicon can drive a homogeneous
wind at $T_\text{eff}=14\,000\,\text{K}$, because silicon dominates the line
force at this temperature.

The dependence of the mass-loss rate on the metallicity has significant
consequences for chemically peculiar stars. The abundance of elements, most
notably that of silicon and iron, varies with location on the stellar surface
\citep[e.g.,][]{choch,leh2}. Consequently, the wind mass flux is variable across
the surface.
For helium-rich stars,
where the abundances of silicon and helium are typically anticorellated, this
means that the wind blowing from helium-rich regions is weaker than that from
helium-poor ones. In magnetic stars the tilt of the flow under the
influence of the magnetic field \citep{owoudan} also influences the mass-loss
rate.

We also provide the wind terminal velocities $v_\infty$ derived from our
models (see Table~\ref{hpvys}). The terminal velocities are sensitive to the
ionization structure of the outer wind, possibly leading to the scatter of the
observed $v_\infty/v_\text{esc}$ ratio in O stars \citep{lsl}. This effect
may explain some variations of $v_\infty$ seen in Table~\ref{hpvys}.
Since in our case the winds are driven mostly by optically thick lines, the
corresponding value of the effective line force parameter $\alpha\rightarrow1$
indicates from simple scaling $v_\infty\sim\alpha/(1-\alpha)Z$ \citep{pusle}
an even stronger sensitivity of $v_\infty$ to the abundance. However, in many cases the
wind may not reach this terminal velocity, either as a result of an inefficient
X-ray cooling or because of multicomponent effects (see Sect.~\ref{modelsi}). This
is also the reason why we argue that the discussion of terminal velocities that
are lower than the escape speed $v_\text{esc}$ would be very premature in some cases
(see~Table~\ref{bhvezpar}).

\begin{figure}[t]
\centering \resizebox{\hsize}{!}{\includegraphics{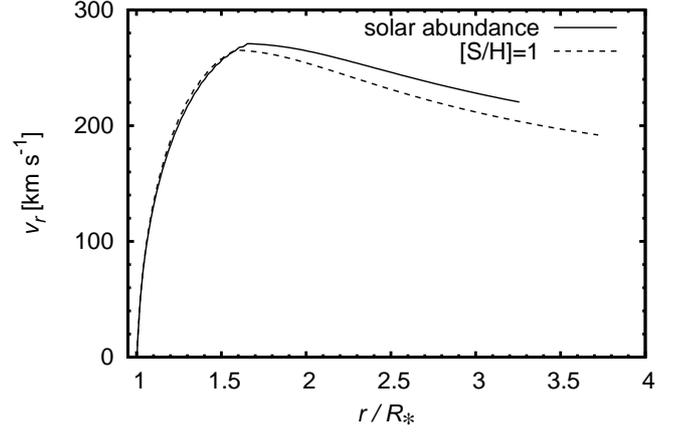}}
\caption{Dependence of the radial wind velocity on radius in selected T16
wind models with a kink for a model with solar chemical composition (solid line) and
with [S/H]=1 (dashed line).}
\label{t16klicka}
\end{figure}

In some cases the radiative force in the outer wind parts is too weak
to maintain an accelerated wind flow. In this case a kink in the velocity
profile occurs \citep{feslop} and wind velocity decreases in the outer parts of
the wind (see Fig.~\ref{t16klicka}). For these the terminal velocity in
Table~\ref{hpvys} corresponds to the highest wind velocity and is given in
parentheses. This typically occurs in the wind models T16. For these
the wind velocity is lower than the escape velocity $v_\text{esc}$ (see
Table~\ref{bhvezpar}), and the wind may never leave the star. We note that our
CMF procedure allows calculating the radiative force in an accelerating wind only,
consequently, the models in Fig.~\ref{t16klicka} were derived assuming a Sobolev
line force with CMF correction.

\section{Wind limit at $\boldsymbol T_\text{eff}\approx15\,000\,\text{K}$}

We were unable to find a wind solution for stars with
$T_\text{eff}<16\,000\,\text{K}$ (except for the silicon-rich ones). This
indicates that these stars have no wind. Therefore, we
provide additional tests for these stars and compare the radiative force with
the gravity. The winds are only possible
when the magnitude of the radiative acceleration $g^\text{rad}$ is larger than
the magnitude of the gravity acceleration $g$,
\begin{equation}
\label{trebova}
g^\text{rad}>g.
\end{equation}
To enable the calculation of the radiative force even  
when there is no wind, the hydrodynamical variables (velocity, density,
and temperature) were kept fixed during the tests.
The velocity structure of our test models is given by an artificial velocity law
\begin{equation}
v(r)=10^{-3}\sqrt{\frac{5}{3}\frac{kT_{\text{eff}}}{m_\text{H}}}+
2 \times 10^{8}\,\text{cm\,s}^{-1}\frac{r-R_*}{R_*},
\end{equation}
where $T_\text{eff}$ is the stellar effective temperature, $R_*$ the stellar
radius, and $m_\text{H}$ the mass of the hydrogen atom. The selection of such
a specific velocity law does not significantly influence our results. The density
structure is obtained from the equation of continuity. In these test models we
assumed a constant wind temperature $\frac{2}{3}T_\text{eff}$, and the electron density
was consistently calculated from the ionization balance. Since for these wind
tests we did not solve the equation of motion, it was necessary to specify the wind
mass-loss rate for which the wind existence was tested. For each set of stellar
parameters these mass-loss rates are
$10^{-12}\,\text{M}_\odot\,\text{year}^{-1}$,
$10^{-13}\,\text{M}_\odot\,\text{year}^{-1}$, 
$10^{-14}\,\text{M}_\odot\,\text{year}^{-1}$, and
$10^{-15}\,\text{M}_\odot\,\text{year}^{-1}$. 

The wind tests support our expectations from the previous section. The wind is
possible for stars with $T_\text{eff}\geq16\,000\,\text{K}$, whereas the
homogeneous winds exist only for silicon-rich stars with
$T_\text{eff}=14\,000\,\text{K}$. From this we conclude that main-sequence stars
with $T_\text{eff}\lesssim15\,000\,\text{K}$ without an enhanced silicon abundance
have no homogeneous winds (i.e., winds composed of hydrogen, helium, and
heavier elements).

We failed to find a wind solution for the T16 model with overabundant iron
$\text{[Fe/H]}=2$. The test using the wind condition Eq.~\eqref{trebova} showed
that the radiative force is unable to drive wind close to the star in a slightly
supersonic region as a result of strong flux redistribution.  We also assumed
that there is no homogeneous wind in this case.

\citet{abbobla} distinguished two wind limits in the H-R diagram. In stars above the
static limit the wind can be self-initiated, that is, the radiative force is
so strong that a static atmosphere is not possible.
All these stars are expected to have winds. In stars
below the static limit and above the wind limit the wind can be maintained if it
already exists. These stars may or may not have winds, depending on initial
conditions. The results of model atmosphere calculations show
that all model stars lie below the static limit. The derived threshold at
$T_\text{eff}\approx15\,000\,\text{K}$ corresponds to the wind limit. In
main-sequence stars our detected wind limit is higher by roughly $2000\,\text{K}$
than that derived by \citet{abbobla}, most likely as a result of more
realistic model atmosphere fluxes. Stars below the wind limit may still have a
purely metallic wind \citep{babela}.

\section{Observational consequences}

\subsection{Comparison of UV line profiles}

The comparison of observed and predicted wind line profiles would be the most
natural test of the derived mass-loss rates. However, this can be complicated
in the domain of main-sequence B stars studied here. In addition to the
clumping problem in more luminous hot stars 
\citep[e.g.,][]{sund,clres1} the weak-wind problem
may also complicate the situation in main-sequence B stars. Consequently, we
here also discuss indirect mass-loss rate indicators and postpone a detailed
study of wind line profiles to a subsequent study.

We note that some studies concentrated on a detailed description of the line
profiles that might be contributed to the winds, instead of on the wind
parameter determination. In some cases this might be a safer option, because
resonance lines may originate both in the wind and in the atmosphere. Detailed
atmosphere models are necessary to judge whether the line originates in the wind
or in the atmosphere \citep{huhas}. \citet{bridlozad} concluded that \ion{Si}{iv} absorption
line profiles are present up to a spectral type of B8
($\Teff=11\,600\,\text{K}$), while he found that \ion{Si}{iv} edge velocities
and equivalent widths do not vary significantly with spectral type for stars
later than about B5 ($\Teff=15\,500\,\text{K}$, see Figs.~4 and 10 therein).
This roughly agrees with the wind limit at
$\Teff\approx15\,000\,\text{K}$ detected by us. The \ion{C}{iv} lines are visible up to a
spectral type of B3 ($\Teff=19\,100$) in standard (i.e., non-Be) stars, in
agreement with their negligible contribution to the radiative force in these
stars (see Table~\ref{carytab}). The observed variations of the \ion{Si}{iv} (or
\ion{C}{iv})  resonance line profiles may be tentatively explained if in 
stars of earlier spectral type than B5 (or B3 in the case of \ion{C}{iv}
lines) we observe absorption caused by the wind, while in the latter ones the
absorption in \ion{Si}{iv} lines originates in the atmosphere.


\subsection{Mass-loss rates of magnetic B stars}

\citet{oskibp} derived the mass-loss rate for magnetic B-type stars from the
ultraviolet line profiles taking into account the X-ray emission. This could
be crucial in the B-star domain \citep[similar to late O dwarfs,][]{martclump}
because of the strong influence of X-rays on the wind ionization equilibrium.
\citet{oskibp} provided two mass-loss rate estimates derived from \ion{C}{iv} and
\ion{Si}{iv} lines. These estimates differ by up to 1~dex, probably as a result
of uncertain wind ionization state.

\begin{table}
\caption{Comparison of mass-loss rates derived by \citet{oskibp} from
\ion{Si}{iv} lines
and the rates predicted from the stellar effective temperature using
Eq.~\eqref{metuje}.}
\centering
\label{maghvezpar}
\begin{tabular}{cccc}
\hline
Star & $\Teff$ $[\text{K}]$ & \multicolumn{2}{c}{$\log(\dot M / 1\,M_\odot\,\text{year}^{-1})$}\\
&&\citeauthor{oskibp}& prediction Eq.~\eqref{metuje}\\
\hline
\object{$\tau$ Sco}      & $30\,700$ & -8.6 & -8.6 \\
\object{$\beta$ Cep}      & $25\,100$ & -9.1 & -9.2 \\
\object{$\xi^1$ξ CMa}    & $27\,000$ & -10  & -8.9 \\
\object{V2052 Oph}  & $23\,000$ & -9.7 & -9.6 \\
\object{$\zeta$ Cas}      & $20\,900$ & -9.7 & -10.2 \\
\hline
\end{tabular}
\end{table} 

Here we selected the mass-loss rates derived from the \ion{Si}{iv} lines,
because these
lines always give a higher mass-loss rate. The weaker \ion{C}{iv} lines can
be explained for instance by a lower ionization fraction of \ion{C}{iv} than derived in the
models. We note that the
opposite (lower mass-loss rate with stronger \ion{Si}{iv} lines) would be more difficult to explain.
In Table~\ref{maghvezpar} we compare these mass-loss rates derived from
observations with predicted rates derived using Eq.~\eqref{metuje} for the effective
temperature given in \citet[see also Table~\ref{maghvezpar}]{oskibp} for the studied
magnetic stars. The mass-loss rates derived from observations and the predicted
rates differ in most cases by no more than 0.1\,dex, consequently, we conclude that
there in this case observation and theory agree well.

The H$\alpha$ emission from the material trapped in the corotating
magnetosphere is typically observable for the stars with effective temperature
higher than about $16\,000\,\text{K}$ \citep{zboril,malykor}. The magnetosphere
is filled by the stellar wind, therefore this anticipated effective
temperature limit for stars with magnetospheric H$\alpha$ emission may be
explained by the disappearance of homogeneous wind at
$T_\text{eff}\approx15\,000\,\text{K}$.

\subsection{Rotational braking of Bp stars}
\label{brzdy}

The helium-rich star \object{$\sigma$ Ori E} shows rotational braking
\citep{town} that can be explained to be the result of angular momentum loss by a
magnetized wind. The spin-down time depends on the stellar parameters, polar
magnetic field strength $B_\text{p}$, and on the wind parameters,
$\tau_\text{spin}\sim M(v_\infty/\dot M)^{1/2}/(B_\text{p}R_*)$ \citep{brzdud}.

For $\sigma$ Ori E we adopt the stellar parameters from \citet[$\Teff=22\,500\,\text{K}$,
$M=8.9\,M_\odot$, $R_*=5.3\,R_\odot$]{huhegr}. For this star we derive from Eq.~\eqref{metuje}
corrected for the assumed radius via $\dot M\sim L^2\sim R^4$ \citep{fosfor}
the mass-loss rate $\dot M=5.5\times10^{-10}\,M_\odot\,\text{year}^{-1}$.
This is a significantly lower value than that derived by \citet{kkiv} as a result
of the use of model atmospheres with metal opacity and CMF line force in the
wind models.

With the polar field strength $B_\text{p}=9.6\,\text{kG}$, moment of inertia
constant $k=0.05$ \citep{mema}, and the terminal velocity
$v_\infty=1690\,\text{km}\,\text{s}^{-1}$ (see Table~\ref{hpvys}), we derive from
Eq.~(25) of \citet{brzdud} a spin-down time of $1.7\,\text{Myr}$. This agrees
well with the value of $1.34\,\text{Myr}$ derived from photometry \citep{town}.

\begin{figure}[t]
\centering \resizebox{\hsize}{!}{\includegraphics{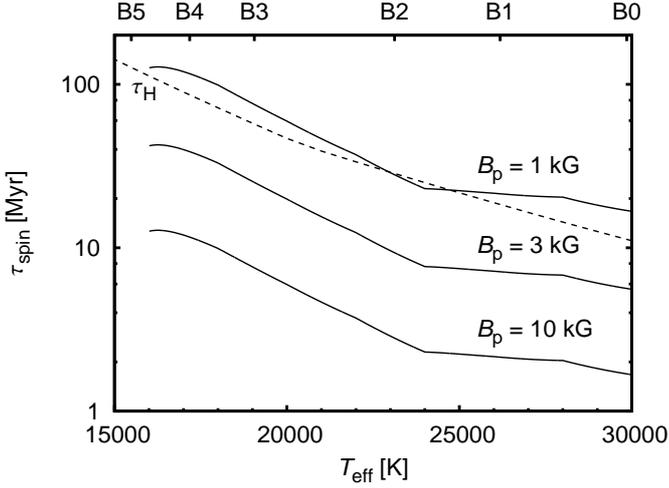}}
\caption{Spin-down time as a result of angular momentum loss via magnetized
stellar wind as a function of effective temperature for different polar field
strengths. Overplotted is the main-sequence lifetime $\tau_\text{H}$ derived
from the non-rotating models of \cite{sylsit}.}
\label{spindota}
\end{figure}

The rotation braking as the result of angular-momentum loss via the magnetized
wind is effective in stars with a wind and a magnetic field. In Fig.~\ref{spindota}
we plot the dependence of the spin-down time \citep{brzdud} for the solar
abundance models studied here for three different values of the polar strength
of the dipolar field. The spin-down time is nearly constant for stars hotter
than about $\Teff=23\,000\,\text{K}$. At this temperature range the decrease of
the mass-loss rate with decreasing temperature is compensated by the decrease of
the $M/R_*$ ratio and $v_\infty$. At lower effective temperatures the spin-down time
significantly increases with decreasing temperature as a result of the
decreasing mass-loss rate.

With long-term observation series of individual stars the rotation braking
with the spin-down time up to roughly 10\,Myr can be detected \citep{zmenper2}.
According to Fig.~\ref{spindota} this means that rotational braking is detectable in
stars with effective temperature $\Teff\gtrsim22\,000\,\text{K}$ and a
medium-strong field, while in cooler stars only a very strong magnetic field causes
detectable period changes. For stars with a polar field intensity stronger than
about $2\,\text{kG}$ the spin-down time is shorter than the main-sequence
lifetime (see Fig.~\ref{spindota}). Consequently, the population of stars with
strong magnetic field should on average show slower rotation than stars with a
weaker field.

These results can be compared with wind-braking in nonmagnetic stars. The
time derivative of the stellar angular momentum in the case of rotational
braking of a uniformly rotating star can be expressed as $\dot J=\eta
MR^2\dot\Omega$, where $\eta$ is a dimensionless constant given by the mass
distribution in the star, and $\dot\Omega$ is the time derivative of the angular
velocity $\Omega$. With angular-momentum loss via a nonmagnetized wind $\dot
J=\frac{2}{3}\dot MR^2\Omega$ (here $\frac{2}{3}$ comes from 
integrating over the spherical surface) the spin-down time is
$\tau_\text{wind}\approx\Omega/\dot\Omega=\frac{3}{2}\eta M /\dot M$. With  $\eta=0.05$
(which is appropriate for a star of mass $M=9~{M}_\odot$, \citealt{mema}), the
spin-down time is higher by at least one order of magnitude than the
main-sequence lifetime. Consequently, rotational braking by a nonmagnetized wind
is negligible in main-sequence B stars.

\section{Discussion}

\subsection{Model simplifications}
\label{modelsi}

Many effects are neglected in our models. The X-ray
emission strongly influences the ionization equilibrium at low effective
temperatures \citep{martclump}. However, our models show that the X-ray
emission does not influence the radiative force close to the star where the
mass-loss rate is determined \citep{nlteiii}. This is supported by
\citet{oskibp}, who showed that the observed level of X-ray emission does not
lead to this decrease of the radiative force and mass-loss rates that would
explain the problem with too weak wind line profiles in low-luminosity stars
\citep[weak-wind problem, e.g.,][]{bourak,martclump}.

Our predicted wind mass-loss rates may be affected by a weak-wind problem.
However, this is probably not the case, because the weak-wind problem
may be caused by a too long cooling length in radiative shocks in the wind
(\citealt{luciebila}, \citealt{cobecru}, \citealt{nlteiii},
\citealt{lucyjakomy}). Such shocks occur well above the critical point
\citep{ocr,felpulpal}, and consequently do not influence the mass-loss rate.
This would mean that the models predict correct mass-loss rates, but the
mass-loss rates derived from the line profiles are too low as a result of the
X-ray overionization. However, this explanation still lacks a detailed proof
using time-dependent hydrodynamical models that take shock heating into account
\citep[e.g.,][]{felpulpal}. Spectra synthesized using such time-dependent models
then should show weak-wind lines that can be compared with the observed lines as
a final test of this scenario. 

Stellar winds of hot stars have a multicomponent nature. Mostly heavier ions and
free electrons are accelerated as a result of absorption of stellar radiation.
The momentum acquired by these components is transferred
to hydrogen and helium via Coulomb collisions. The wind can be treated as a one-component flow at high
wind densities, while at low wind densities typical for main-sequence B star
winds the effects connected with the multicomponent flow are significant.
However, either frictional heating \citep{treni,kkiv} or the decoupling of wind
components \citep{op,ufo} typically occur at velocities higher than the critical
one, therefore they do not affect the wind mass-loss rate,
but may affect the terminal velocity.

\begin{figure}[t]
\centering \resizebox{\hsize}{!}{\includegraphics{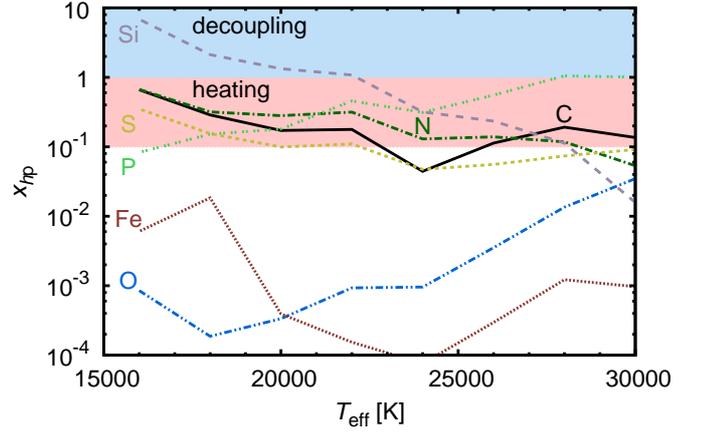}}
\caption{Variations of the maximum relative velocity difference of
individual elements and protons as a function of stellar effective temperature
in models with solar chemical composition.}
\label{xpi}
\end{figure}

The importance of the multicomponent effects is given by a relative
velocity difference $x_{h\text{p}}$ \citep[Eq.~(18) in][]{nlteii} between a
given element and protons. For $x_{h\text{p}}\lesssim0.1$ the multicomponent
effects are unimportant, for $x_{h\text{p}}\gtrsim0.1$ the frictional heating
typically influences the wind temperature, and for $x_{h\text{p}}\gtrsim1$ 
decoupling of components is possible. We calculated the relative velocity
difference following \citet{nlteii} for all considered elements in all models. The
relative velocity difference between wind components increases with radius as a
result of the decreaseing wind density. In all models presented here (except
for the T14 silicon-rich models) the relative velocity difference is lower than 1 at the
critical point, therefore the decoupling does not affect the mass-loss rate,
which is determined at the critical point.

In some models the multicomponent effects may influence the terminal velocity.
This is documented in Fig.~\ref{xpi}, where we plot the maximum relative
velocity difference $x_{h\text{p}}$ between a given element and protons found in
the wind. For the hottest stars $T_\text{eff}\gtrsim28\,000\,\text{K}$
phosphorus may decouple from the wind. However, this probably does not
significantly affect the wind, because the contribution of phosphorus to the
radiative force is negligible. For stars with
$T_\text{eff}\gtrsim24\,000\,\text{K}$ the multicomponent nature of the flow
leads to additional heating due to carbon, nitrogen, and silicon. For cooler
stars with $T_\text{eff}\lesssim22\,000\,\text{K}$ the decoupling of elements
may occur in the outer parts of the wind. For these stars the terminal
velocities in Table~\ref{hpvys} needs to be corrected for multicomponent effects.

In O stars the winds are typically relatively dense and the stars themselves are
luminous, which makes the typical time needed for the ionization or
recombination very short, in most cases much shorter than the typical
flow-time. As a result, the advection term in NLTE equations can be safely
neglected \citep{lamo}. This is no longer true in B stars \citep{oskibp}
with their weaker radiation field and thinner winds. We compared the ionization and
recombination rates of our model ions with the inverse of the flow time $v/r$,
where $v$ is the radial velocity. From our models it follows that for stars with
$T_\text{eff}\gtrsim20\,000\,\text{K}$ the advection term for dominant ions is
negligible in the inner parts of the wind where the mass-loss rate is
determined. The relative unimportance of advection term for the dominant
ionization states (that drive the wind) follows from the fact that the
ionization from lower ionization state is fast, while the corresponding
recombination is slow, and the opposite is true for the higher ionization state.
On the other hand, in these stars the advection term may affect the population
of minor ionization states and even also of dominant ones in the outer wind
regions, where the lines used as observational indicators originate
\citep{martclump}. For stars cooler than about $T_\text{eff}=20\,000\,\text{K}$
the radiative processes become slower than the wind radial expansion,
and in turn the advection term may become important even for the dominant ions
close to the critical point. For these stars the ionization equilibrium becomes
partially frozen in the flow and the advection term may affect the wind mass-loss
rate.

\subsection{Wind initiation}

The calculated model atmospheres show that the radiative
force in the {\em static} model atmospheres in main-sequence B stars does not
overcome the gravity (up to $\tau_\text{ross}\approx10^{-7}$). This means that
all model stars lie below the static limit introduced by \citet{abbobla}.
We have shown that two different solutions are possible, that is, a static
atmosphere and an extended atmosphere with wind. We assumed that the
wind was already initiated.
It is 
unclear how the winds are initiated in this case, or even if they are initiated at
all. There are several possibilities how the winds might be initiated.
The radiative force on individual heavier elements overcomes the gravity in
all models. This means that a metallic wind can be initiated first from
the uppermost parts of the stellar atmosphere, which
subsequently also drives hydrogen and helium out from the stellar atmosphere.
The other possibility is that a random perturbation in the stellar
atmosphere de-shadows the line from the photospheric line profile, leading to the
initiation of an outflow. Winds may also be initiated in the pre-main-sequence
phase, when the star has a lower surface gravity. 
The wind initiation is an interesting problem on its own, which deserves
further study.

\subsection{Filling the magnetosphere of $\sigma$ Ori E}

There are few Bp stars (including \object{$\sigma$ Ori E}) that show light
variability as a result of light absorption in the circumstellar material
trapped thanks to the interplay between magnetic field and centrifugal force.
The amount of the light variability enables us to estimate the mass of the
magnetosphere and (with the wind mass-loss rate) the time needed to fill the
clouds.

The amplitude of the $\sigma$~Ori~E light variability is about 0.2\,mag,
which means that about 20\% of the stellar flux is absorbed by the clouds, implying
the minimum optical depth of the clouds is $\tau=0.2$. If the magnetosphere has the
same opacity as the atmosphere, then the mass of the magnetosphere can be
estimated as the column mass of the atmosphere $m(\tau=0.2)$ at the optical
depth $\tau=0.2$ multiplied by the stellar surface area. From the model
atmosphere of $\sigma$~Ori~E we derive
$m(\tau=0.2)=0.1\,\text{g}\,\text{cm}^{-2}$, giving the total mass of
$\sigma$~Ori~E magnetosphere $M_\text{m}=4\pi
R_*^2m(\tau=0.2)\approx10^{-10}\,\text{M}_\odot$. This should be regarded as a
minimum mass, because the material could be optically thick, which would mean
that the optical depth were significantly higher.

Another estimate of the mass of the magnetosphere can be derived by assuming that
the light scattering on free electrons dominates the opacity. In the ionized
medium the opacity cannot be lower than that given by the light scattering on
free electrons. Then the absorbing column mass $m_\text{e}=\tau
m_\text{H}/\sigma_\text{Th}$, where $\sigma_\text{Th}$ is the Thomson scattering
cross-section, and $m_\text{H}$ the hydrogen mass. This gives another estimate of
the magnetospheric mass, $M_\text{m}\approx4\pi R_*^2
m_\text{e}=4\times10^{-10}\,\text{M}_\odot$. 

From the mass-loss rate of $\sigma$~Ori~E given in Sect.~\ref{brzdy} we can
conclude that the shortest typical time needed to fill the magnetosphere of a star
similar to $\sigma$~Ori~E is about one year. This is shorter by two orders of magnitude
than the breakout time of the magnetosphere estimated by \citet{towo}.
From this we can conclude that many other Bp stars should show clear signatures
of the matter trapped in the magnetosphere. However, this is not the case
because such stars are very rare. Therefore we can infer that there has to
be some as yet unknown process that incessantly removes the material deposited by
the winds from the magnetospheres of most Bp stars
\citep{nemamelaby}.

\subsection{How are wind and peculiarity connected?}

Some B and A stars show chemical peculiarity caused by a radiative diffusion
\citep[see, e.g.,][for a review]{rompreh}. The slow motion caused by the wind in
the atmosphere interacts with diffusion and may modify the observed abundance
anomalies. The interplay between winds and peculiarity depends on the wind
mass-loss rate, and the result is therefore different for stars with different
effective temperatures or spectral types.

In helium-rich stars a very weak wind with a mass-loss rate of the order of
$10^{-13}-10^{-12}\,M_\odot\,\text{year}^{-1}$ is required to explain the
overabundance of helium \citep{vasam,vadog}. Typically, helium-rich stars
span a temperature range of about $T_\text{eff}=18\,000-23\,000\,\text{K}$
\citep{zboril,HuGr}. For stars in this temperature range the solar abundance
mass-loss-rate predictions give a mass-loss rate higher by one to two orders of magnitude
than that required by a diffusion theory. Such a large mass-loss would
destroy any chemical peculiarity \citep{vimiri}. On the other hand,
helium-rich regions typically are metal-poor \citep{choch,bohacen}, implying a
lower wind mass-loss rate. It seems that the helium overabundance may be
maintained if it already exists, but it is not clear how the atmosphere was able
to
evolve to this state. Possibly the line-driven wind
mass-loss rate increases as the effective temperature increases during star formation \citep{palas},
leading to helium overabundance at the time of appropriate mass-loss rate. 
We also note that in the presence of the magnetic field the surface mass-flux
$\dot m$ is
proportional to the tilt $\theta_B$ of the wind with respect to the local
vertical direction, $\dot m\sim\cos^2\theta_B$ \citep{owoudan}.
This might be another reason
for the decreasing wind mass-loss rate even to values required by the
diffusion theory in stars with overabundant helium.

In chemically peculiar stars with low effective temperatures
($T_\text{eff}\leq15\,000\,\text{K}$) the interplay of peculiarity and winds
with mass-loss rates of the order of
$10^{-14}-10^{-12}\,M_\odot\,\text{year}^{-1}$ should lead to observable
overabundances of oxygen or neon \citep{nedolez}. Oxygen overabundances are not
observed, possibly implying the absence of such winds in these stars. This 
agrees with our finding that stars in this temperature range have no
homogeneous line-driven wind. On the other hand, \citet{mirivi} argued that
some mixing is required to explain the observed abundances in Sirius~A, possibly
with the stellar wind with mass-loss rates
$10^{-13}\,M_\odot\,\text{year}^{-1}$, which is not expected from our models.
These stars could have pure metallic wind with mass-loss rates lower by about three
orders of magnitude \citep{babela}, which is too weak to cause 
substantial mixing. Possibly, the required mixing is caused
by the turbulence.

Despite this qualitative agreement, some open questions remain. For example, the
chemically peculiar star \object{CU Vir} shows pulsed radio emission
\citep{trigilio,kellett}, which requires a wind mass-loss rate
of about $10^{-12}\,{M}_\odot\,\text{year}^{-1}$ \citep{letcuvir}. However,
stars with an effective temperature $T_\text{eff}=13\,000\,\text{K}$
of \object{CU Vir} \citep{kus} can have a purely metallic wind with a mass-loss
rate lower by about three magnitudes \citep{babela}. It is unclear how 
this disagreement can be solved. The wind might originate from silicon-rich
parts of \object{CU Vir} surface, but the predicted mass-loss rate (see
Table~\ref{hpvys}) is still too low by at least one order of magnitude.

Chemically peculiar stars typically display an inhomogeneous surface distribution
of elements with relative elemental abundance differences reaching a few
orders of magnitude \citep[e.g.,][]{briketka,leh2,bohacen}. As a result of the
dependence of the mass-loss rates on abundance (see Table~\ref{hpvys}), the mass
flux from individual surface elements varies with location on the stellar
surface. Moreover, the tilting of the flow in the presence of the magnetic
field \citep{owoudan} and possibly also the Zeeman effect influence the
mass-loss rate. These effects may contribute to the observed rotational
variability of the strength of ultraviolet resonance lines in Bp stars
\citep{shobro}.

\subsection{Winds of fast-rotating stars}

The B star mass-loss rates are very sensitive to the stellar effective
temperature. For fast-rotating stars where the gravity darkening becomes
important this leads to a high ratio of polar to equatorial mass-loss rate and
density. Taking as an example the star \object{HR 7355} that rotates at about $90\,
\%$ of its critical rotational velocity \citep{ritok} with a polar effective
temperature of $19\,800\,\text{K}$ and an equatorial temperature of
$15\,700\,\text{K}$, the
mass-loss rate formula Eq.~\eqref{metuje} yields the polar to equatorial
mass-loss-rate ratio of about 40. Here we neglected the nonradial component of
the radiative force \citep{bezdisko}, but this result still shows that the wind
asymmetries in fast-rotating main-sequence B stars can be enormous.

\section{Conclusions}

We calculated NLTE radiatively driven wind models of main-sequence B stars with
the CMF line force. We provided the wind mass-loss-rate predictions as a function of
the stellar effective temperature. The early-B stars have line driven winds with
mass-loss rates of the order of $10^{-9}\,\text{M}_\odot\,\text{year}^{-1}$. For
the cooler stars the mass-loss rate strongly decreases with their effective
temperature, which means that the solar-metallicity main-sequence B stars with effective
temperatures lower than about $15\,000\,\text{K}$ have no homogeneous
line-driven wind. This general picture qualitatively agrees with observations of
UV resonance lines and with recent B star mass-loss-rate determinations.

Winds of main-sequence B stars are driven mainly by carbon, nitrogen, and
silicon. In the winds of stars with $T_\text{eff}\lesssim22\,000\,\text{K}$ the
silicon-driving dominates.

The mass-loss rates depend on the elemental abundances. For elements that do not
significantly alter the emergent model atmosphere flux the mass-loss rate
increases with increasing abundance. However, for elements that significantly
influence the emergent flux (most notably iron) the mass-loss rate may decrease
with increasing abundance of a given element. The dependence of the radiative
force on the silicon abundance in the coolest stars studied here is so strong
that wind may be possible even below the solar-abundance wind limit at
$T_\text{eff}=15\,000\,\text{K}$.

We discussed the implications of our models for the rotational braking, filling
the magnetosphere of Bp stars, and chemical peculiarity.

\begin{acknowledgements}
This work was supported by grant GA \v{C}R 13-10589S.
\end{acknowledgements}

\end{document}